\documentclass[12pt,preprint]{aastex}
\usepackage{emulateapj5}

\shorttitle{Alternative to Cooling Flows?}
\shortauthors{McCarthy, West, \& Welch} 

\begin{document} 

\title{\bf Non-thermal X-ray Emission: An Alternative to Cluster Cooling Flows?}

\author{Ian G. McCarthy$^{1,2}$, Michael J. West$^{3}$, and Gary A. Welch$^1$}

\affil{$^1$Department of Astronomy \& Physics, Saint Mary's University, Halifax, NS, 
B3H 3C3, Canada}

\affil{$^3$Department of Physics \& Astronomy, University of Hawaii at Hilo,
Hilo, HI, 96720, U.S.A.}                              

\footnotetext[2]{current address: Department of Physics \& Astronomy,  
University of Victoria, Victoria, BC, V8P 1A1, Canada; mccarthy@beluga.phys.uvic.ca}

\begin{abstract} 

We report the results of experiments aimed at reducing the major problem with cooling flow
models of rich cluster X-ray sources: the fact that most of the cooled gas or its products
have not been found.  Here we show that much of the X-ray emission usually 
attributed to cooling flows can, in fact, be modeled by a power-law component which is 
indicative of a source(s) other than thermal bremsstrahlung from the intracluster 
medium.  We find that adequate  simultaneous fits to ROSAT PSPCB and ASCA GIS/SIS 
spectra of the central regions of ten clusters are obtained for two-component models 
that includes a thermal plasma component that is attributable to hot intracluster gas 
and a power-law component that is likely generated by compact sources and/or extended 
non-thermal emission.  For five of the clusters that purportedly have massive cooling 
flows, the best-fit models have power-law components that contribute $\sim$ 30 \% of 
the total flux (0.14 - 10.0 keV) within the central 3 arcminutes.  Because cooling flow 
mass deposition rates are inferred from X-ray fluxes, our finding opens the possibility 
of significantly reducing cooling rates.

\end{abstract}

\keywords{cooling flows ---  X-rays: galaxies --- X-rays: galaxies: clusters}

\section{Introduction}

It is well established that the diffuse X-ray emission from clusters of galaxies implies
the existence of a hot (10$^{7-8}$ K), thermally radiating intracluster medium (ICM).  In
the dense centers of many clusters this gas is expected to cool radiatively in less than 
a Hubble time.  As pressure support decreases the gas will gravitate towards the bottom 
of the cluster potential well, a phenomenon that has come to be known as a ``cooling 
flow'' (cf. Fabian 1994).  Such cooling flows can be detected by direct X-ray imaging, 
where they are seen as a surface brightness enhancement in the X-ray emission from a 
region typically $\sim$ 100 - 200 kpc in radius, and by spectroscopic observations which 
reveal gas at cooler temperatures than the surrounding ICM.  The excess X-ray 
luminosity in the cooling flow region is typically 10$^{42} < L_{cool} < 10^{45}$ ergs 
s$^{-1}$, corresponding to roughly  10 - 50 \% of the total cluster X-ray emission 
(Peres et al. 1998).  The mass deposition rate is $\dot{M} \propto L_{cool}T^{-1}$, where 
$T$ is the gas temperature at r$_{cool}$, the cooling radius.  Estimated cooling flow 
rates in the centers of rich clusters vary from 30 - 300 M$_{\odot}$ yr$^{-1}$ (White, 
Jones, \& Forman 1997) and in some cases reach mass deposition rates of 1000 M$_{\odot}$ 
yr$^{-1}$ or more (Allen 2000). 
 
A long-standing problem with the cooling flow scenario has been the fact that the large
quantities of cool gas expected to accumulate from the flow have rarely been detected.  
Searches for cool gas in atomic or molecular form have generally proven unsuccessful 
(e.g., Heckman et al. 1989; Jaffe 1990; McNamara, Bregman, \& O'Connell 1990; O'Dea et 
al. 1994; Voit \& Donahue 1995; Jaffe \& Bremer 1997; Donahue et al. 2000), with reported 
upper limits two orders of magnitude smaller than the masses implied by X-ray mass 
deposition rates (although see Edge 2001).  While some support for the cooling 
flow 
picture is provided by low energy X-ray spectra of clusters which appear to require 
larger absorption than can be accounted for by foreground gas alone (White et al. 1991; 
Allen et al. 1993; Irwin \& Sarazin 1995), it now appears that the fraction of 
clusters that actually possess such excess absorption may be much less than previously 
thought (Arabadjis \& Bregman 2000, hereafter AB2000; B\"{o}hringer et al. 2001a).  
Furthermore, recent high-resolution X-ray spectra obtained with the XMM-Newton and 
Chandra satellites have failed to show expected emission lines from cooled gas at 
temperatures below a few keV (Fabian et al. 2001; Peterson et al. 2001; Tamura et al. 
2001) and it has also been found that local isothermality matches the spectral data 
of ``cooling flow'' regions better than the continually declining temperature profile 
predicted by the standard cooling flow model (e.g. Molendi \& Pizzolato 2001; 
B\"{o}hringer et al. 2001a).  A variety of mechanisms have been proposed to account 
for the lack of detectable cool gas in cooling flows; these include reheating the gas, 
mixing, differential absorption, efficient conversion of cooling flow gas into low mass 
stars, inhomogeneous metallicity distributions, or disruption of cooling flows by 
recent subcluster mergers (see recent discussions in Fabian et al. 2001 and Peterson et 
al. 2001).  As of yet, it is unclear whether any of these mechanisms are capable of 
resolving the aforementioned problems with the cooling flow model.  

One characteristic that is common to the mechanisms mentioned above is that they have 
mainly been introduced in an ``after the fact'' manner.  That is, they assume the 
observed mass deposition rates are basically correct but that something is happening to 
the gas after (or while) it cools, be it reheating or conversion into low mass stars or 
some other process.  Perhaps a simpler way of addressing the problems of the cooling 
flow model are to look for ways to reduce the mass deposition rates from the beginning.  
One way to achieve this is to reduce the excess emission ($L_{cool}$) associated with 
cooling flows by attributing some (or all) of it to X-ray sources other than thermal 
bremsstrahlung or line emission from the ICM.  But are there sources luminous enough to 
reproduce $L_{cool}$, can they be found within galaxy clusters, and, most importantly, 
can they mimic hot cluster gas?      

Soon after X-ray emission was first detected from galaxy clusters, Katz (1976) proposed
that much of this emission might come from individual compact sources such as those known
to exist in our own Galaxy.  However, the subsequent discovery of iron emission lines in
X-ray spectra (Forman \& Jones 1982, and references therein) clearly demonstrated that
thermal bremsstrahlung from a diffuse intracluster gas is the dominant source of X-ray
emission in clusters, and interest in compact sources quickly waned.  Nevertheless,
compact sources must surely contribute some fraction of the total X-ray emission from
galaxy clusters.  A growing body of evidence indicates that a substantial intergalactic
population of stars, star clusters and stellar remnants is present in the cores of rich
clusters, and some of these objects will be X-ray sources.  For example, Ferguson, Tanvir,
\& von Hippel (1998) showed the existence of intergalactic stars in the Virgo cluster,
and planetary nebulae associated with this population have also been detected (e.g.,
Arnaboldi et al. 1996; Ciardullo et al. 1998).  Similarly, West et al. (1995) hypothesized
the existence of a large population of intergalactic globular clusters in the cores of
galaxy clusters based on observed galaxy-to-galaxy variations in globular cluster
populations.  Furthermore, there is mounting evidence that many cluster galaxies have
been destroyed over a Hubble time, victims of collisions with other galaxies 
or disruption by the mean tidal field of the cluster potential in which they reside
(e.g., Thompson \& Gregory 1993; Lopez-Cruz et al. 1997; Gregg \& West 1998; Moore et
al. 1999; Calcaneo-Rodin et al. 2000).  As galaxies are disrupted they will spill their
contents throughout intracluster space, including X-ray sources such as X-ray binaries,
globular clusters, and black holes.  Such disruption is likely to be most frequent in the
dense cluster core, and hence would be expected to produce a substantial pool of 
intracluster material, including compact X-ray sources, at the cluster center.  Any
significant contribution by these compact sources to the X-ray flux will have to be
taken into account when estimating the properties of ICM, including temperatures,
densities, and cooling flow mass deposition rates.

An intracluster population of compact objects would have a different spectral signature
than thermal emission from intracluster gas.  For instance, the spectra of large spiral 
and elliptical galaxies reveal the presence of hard X-ray emission which is thought to 
be produced by either high mass X-ray binaries or accretion onto massive black holes and 
is adequately represented by power-laws having $\Gamma$ $\sim$ 0.6 - 3.0 (where $F_{\nu} 
\propto \nu^{-\Gamma}$; Colbert \& Mushotzky 1999; Allen, Di Matteo, \& Fabian 2000).  
Luminous AGN, on the other hand, are often fit by power-laws with $\Gamma$ $\sim$ 
1.6 - 2.0 (Nandra et al. 1997).  Indeed, the spectral signatures of such objects have 
already been identified in several X-ray clusters by Allen et al. (2001).  Using ASCA 
data, these authors found very luminous sources with power-law spectral indices 
characteristic of compact sources.

The recent discovery of luminous {\it extended} non-thermal X-ray emission in several 
weak (or non-) cooling flow clusters (Rephaeli, Gruber, \& Blanco 1999; Fusco-Femiano et 
al. 1999; 2000; 2001) suggests that one should also look for such emission in massive 
cooling flow clusters.  At present, the origin of the diffuse non-thermal emission is 
still a matter of some debate.  Suggested mechanisms include inverse-Compton scattering 
of CMB photons up to X-ray energies via a population of highly relativistic electrons, 
and non-thermal bremsstrahlung, likewise, by a population of relativistic electrons 
(Sarazin 2001; Rephaeli 2001).  Those highly relativistic electrons may have been 
accelerated by recent subcluster merger events which shocked the ICM or may have been 
ejected from a central massive black hole.  The discovery of X-ray 'holes' at the  
positions of radio lobes of a central compact object in several massive cooling flow 
clusters (e.g. B\"{o}hringer et al. 1993, McNamara et al. 2000; Fabian et al. 2000) 
certainly lends credence to the second hypothesis.  Recently, Harris et al. (2000) 
(Chandra data) and B\"{o}hringer et al. (2001b) (XMM-Newton data) have even identified 
extended non-thermal emission directly associated with radio jets/lobes in 3C295 and M87, 
respectively. 

The spectral signature of extended non-thermal emission is also different from that of 
thermal emission.  To date, the extended non-thermal components of X-ray spectra of 
clusters have adequately been modeled by power-laws having $\Gamma \sim 0.5 - 6.0$ 
(Fusco-Femiano et al. 1999; 2000; 2001).  As with luminous compact sources, a proper 
deduction of cluster properties should explicitly include the effects of extended 
non-thermal emission in the analysis.

In the present work we assess the case for a power-law component to the X-ray spectra of
cluster centers, in particular those with cooling flows.  We show that simultaneous
fits of ROSAT PSPCB and ASCA GIS/SIS spectra, which have been used to infer the
presence of cooling flows in many clusters to date, are in fact consistent with a
significant contribution from compact sources and/or extended non-thermal emission.  
Using ROSAT HRI data, we have placed limits on the contribution of any bright point 
sources within the central regions of these clusters and conclude that the power-law 
components are most likely produced by either a collection of many low-luminosity 
compact sources (with $L_X \lesssim 5 \times 10^{42}$ ergs s$^{-1}$) or by diffuse 
non-thermal emission or by some combination of the two.  We shall assume $H_{o}$ = 50 km 
s$^{-1}$ Mpc$^{-1}$ and $q_o = 0$ throughout.

\section{Data Selection and Reduction}

Ten well-studied X-ray luminous galaxy clusters (Table 1) were chosen to span a wide
range of cooling flow mass deposition rates.  Five of these have essentially no cooling
flow (NCF), while the other five cooling flow (CF) clusters have  calculated mass
deposition rates ranging from 189 to 645 solar masses per year.  Observations by the
ROSAT Position Sensitive Proportional Counter (PSPCB) and High Resolution Imager (HRI) 
and the ASCA Gas Imaging Spectrometer (GIS) and Solid-state Imaging Spectrometer (SIS)  
were obtained from the HEASARC public archives and reduced using FTOOLS v5.01.  The 
combination of ROSAT and ASCA data provides good sensitivity to both soft and hard X-ray 
emission, spanning the total energy range 0.14 - 10.0 keV, and should provide a strong 
test of power-law components in the X-ray spectra of clusters.     

\subsection{ROSAT observations}

The ROSAT PSPCB data were corrected for spatial and temporal gain fluctuations in
the detector.  X-ray spectra, spanning the energy range 0.14 - 2.04 keV, of all ten
clusters were extracted from circular regions centered on the peak of the emission with a
constant extraction radius (R) of 3 arcminutes.  Background spectra were selected from 
source-free circular regions of radius R.  All spectra were corrected for the effects of 
cosmic rays in the detector (Snowden et al. 1992), vignetting, spatial variations in 
detector efficiency, and deadtime.  The background-subtracted spectra were then grouped  
to ensure a minimum of 20 counts per radial bin so that reliable $\chi^{2}$ statistics 
could be computed.

The ROSAT HRI data was reduced in a manner similar to the PSPCB data.  The HRI data was 
used to place limits on the contribution of any point sources within the central 
regions of these clusters.  Unfortunately, the HRI lacks spectral resolution and was 
not used in the main spectral analysis.  A description of the analysis of HRI data is 
given in Section 5.  No HRI data were available for A2063 or A3532.  

\subsection{ASCA observations}

The screened events from ``revision 2'' processing of ASCA GIS/SIS data were
used.  Additional cleaning procedures (discussed in the ASCA ABC Guide) were 
implemented, including gain and deadtime corrections.  For the GIS detectors (GIS2 and 
GIS3), X-ray spectra spanning the energy range 0.7 - 10.0 keV were extracted from the 
same region of sky as that of the ROSAT PSPCB source spectra.  For the SIS detectors 
(SIS0 and SIS1), X-ray spectra spanning the range 2.0 - 10.0 keV were extracted, thus 
avoiding complications with the degrading efficiency of the SIS detectors at low 
energies.  Background spectra for both the SIS and GIS detectors were extracted from the 
``blank sky'' observations made during the performance testing stage of the mission.  
Spectra were extracted from the same regions of the detectors as that of 
the source spectra (thus minimizing position-dependent effects associated with 
background subtraction).  The background-subtracted spectra were also grouped (or 
binned) to ensure $\chi^{2}$ statistics could be used.  For the clusters A2142 and A3667 
SIS1 data were ignored because of an excess of hot and flickering pixels.  


\section{Spectral Analysis}

Reduced ROSAT PSPCB and ASCA GIS/SIS spectra were fit simultaneously using models within 
the software package XSPEC v11 (Arnaud 1996).  Each cluster was fit with a variety of 
models based on MEKAL thermal plasma routine (Mewe \& Gronenschild 1985; Kaastra 1992; 
Liedahl et al. 1995), and/or redshifted power-laws.  In addition, we used the popular 
cooling flow model comprising a thermal plasma (MEKAL) and cooling flow component 
implemented by the MKCFLOW routine within XSPEC (Mushotzky \& Szymkowiak 1988).  Fits 
were obtained by varying the temperature ($kT$) and metallicity ($Z$) of thermal plasma 
component(s) and by varying the photon index ($\Gamma$) of any power-law components.  
Photoelectric absorption (Balucinska-Church \& McCammon 1992) was applied to the 
spectral models through two separate methods: i) by applying only one absorption 
component to the entire spectral model and fixing the HI column density ($N_H$) at the 
known Galactic value (see Table 1) and then letting it vary to see if the fit could be 
improved; ii) by applying one absorption component to the entire spectral model (with 
$N_H$ fixed at the Galactic value) and a separate redshifted variable absorption 
component ($N_{H,PL}$) to any power-law components via the ZPHABS model in XSPEC.  The 
second method allows for {\it intrinsic} absorption to be associated with any power-law 
components.  Spectra of nearby high mass X-ray binaries and accreting massive black 
holes provide strong evidence for heavy intrinsic absorption, with $N_H \sim 10^{21-23}$ 
cm$^{-2}$ above the Galactic column (Colbert \& Mushotzky 1999; Iwasawa et al. 2000; 
Akylas et al. 2001).  We therefore regard the addition of a free intrinsic absorption 
parameter as an important feature of our spectral analysis.    

Henceforth we will use TP, PL, and CF to refer, respectively, to thermal plasma (MEKAL), 
power-law, and MKCFLOW components.  For example, in this notation TPPL identifies a model 
having thermal plasma and power-law components.  Goodness of fit was judged by the 
reduced chi-square ($\chi^{2}_{r}$) and by visual inspection of residual plots.  
When additional components were added to a spectral model, for example adding a 
power-law component to a thermal plasma component to make TPPL, the corresponding 
F-statistic was calculated in order to judge whether the addition was reasonable. 
Unless stated otherwise, the derived relative contributions of each of the components to 
the total flux refer to the energy range 0.14 - 10.0 keV stated, whereas the 
luminosities presented refer to the ASCA GIS band only (0.7 - 10.0 keV), since the 
spectra were fit simultaneously and not combined.  All luminosities presented were 
corrected for Galactic absorption only.   

To check our data reduction procedures, spectra from each of the detectors were also fit 
individually and no significant deviations were found.  We have also attempted to 
duplicate published results by Arabadjis \& Bregman (2000) for three clusters derived 
using the same data (ROSAT PSPCB) and the same software.  The results, which refer to 
the inner 3 arcminute radius, are presented in Table 2.  Here, the hotter TP component 
is fixed at the White, Jones, \& Forman (1997) value, $Z = 0.5 Z_{\odot}$, $N_H$ is 
fixed at the Stark et al. (1992) value, and the energy range is 0.14 - 2.4 keV and 
0.14 - 2.04 keV for AB2000 and our study, respectively.  The comparison shows excellent 
agreement, with derived temperatures and mass deposition rates having the same values to 
within one standard deviation.  


\section{Results}

\subsection{Non-Cooling Flow Clusters}

We find that single-temperature TP models provide acceptable fits to the spectra of 
A119, A2063, A3266, A3532, and A3667 (1.00 $\leq$ $\chi^{2}_{r}$ $\leq$ 1.17) with 
temperatures in very good agreement with previous studies (White, Jones, \& Forman 1997; 
Markevitch et al. 1998).  Significant improvement is not usually obtained 
by varying the hydrogen column density from the known Galactic value reported by 
Dickey \& Lockman (1990) or Stark et al. (1992).  The derived metallicities range from 
$0.2 \lesssim Z \lesssim 0.4$ which is typical of the intracluster medium (Allen \& 
Fabian 1997).  

In general, both pure PL models with or without variable intrinsic absorption do not 
provide acceptable fits to the NCF cluster spectra with $\chi_r^2$ $\gtrsim$ 1.4.  This 
result is not unexpected since it leaves no room for thermal emission from the ICM.  
With intrinsic absorption included, an acceptable fit is obtained for A3532 with  
$\chi_r^2 \approx 1.05$, however.  This cluster is located at the center of the Shapley 
Concentration and is thought to be undergoing a merger with A3528.  The diffuse  
non-thermal emission generated from such a merger might possibly account for the ability 
of the simple PL model to fit the spectrum of A3532.  Unfortunately, no HRI data exists 
for this cluster so we cannot place limits on the contribution of any point sources 
within  the central 3 arcminutes. 
      
\vskip0.1in
 
{\epsscale{1.0}
\plotone{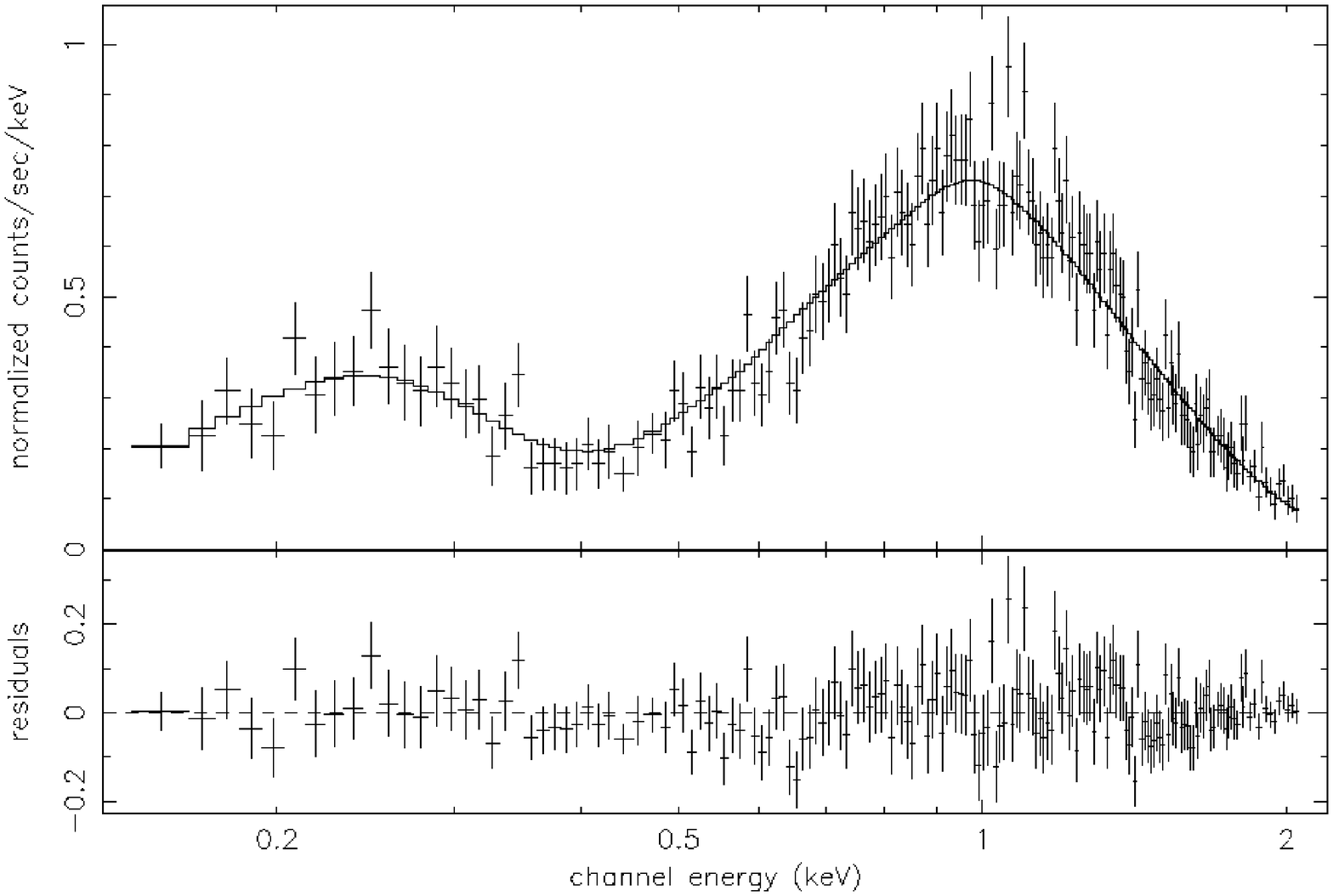}
{Fig.1. \footnotesize
The best fitting TPPL model on the spectrum of A3667.  The
power-law component has a photon index of 1.30 and contributes $\approx$ 22
\% of the flux over the range 0.14 - 10.0 keV.  Only the ROSAT PSPCB data are
presented here.
}}
 
\vskip0.1in

In the best fitting TPPL models, the PL component typically contributes 1 - 14\% (mean 
of 7 \%) when absorbed by the Galactic column only or 10 - 25 \% (mean of 20 \%) when 
intrinsic absorption is included (excluding A3532 which has an anomalously large PL 
component).  We note that in virtually every case the inclusion of intrinsic absorption 
noticeably improves the quality of the fit, making it superior to even the simple TP 
models (see Table 3).  A mean F-statistic of $\approx 10$ ($\gtrsim$ 95 \% 
confidence) was found when the PL component with intrinsic absorption was included in 
the fit indicating that its addition is, indeed, reasonable.  The corresponding  
luminosities of intrinsically absorbed PL components range from $L_{X,0.7-10.0}\approx 
5 \times 10^{42} - 7 \times 10^{43}$ erg s$^{-1}$ with a mean value of $L_{X, 0.7-10.0} 
\approx 2.2 \times 10^{43}$ erg s$^{-1}$ (corrected for Galactic absorption only).  
Photon indices for these components are typical of compact accretion-driven X-ray 
sources and extended non-thermal emission.  Interestingly enough, the amount intrinsic 
absorption ranges from $1.8 \times 10^{21} \leq N_{H,PL} \leq 1.4 \times 10^{22}$ 
cm$^{-2}$ which is typical of optically thick compact sources.  We also note that the 
temperature of the ICM tends to decrease significantly when a power-law component is 
included.  This trend was previously reported by Allen et al. (2001) when including a 
power-law component in their spectral models.  Figure 1 presents the best fit TPPL 
model to the spectrum of A3667.  For clarity, only the ROSAT PSPCB data are plotted.


As expected, standard cooling flow (TPCF) models of our non-cooling flow sample yield 
best fits with $\dot{M}$ = 0 M$_{\odot}$ yr$^{-1}$ to within one sigma.  The mass
deposition rate for the modest cooling flow cluster A2063, however, is 25 $\pm$ 12 
M$_{\odot}$ yr$^{-1}$.  The cooler component in our TPTP models typically contributes 
zero flux to within one sigma ($\sim$ 10 \% for A2063 \& A3667).  Neither of these models 
provided fits significantly better than those of the standard TP models.  

In summary, we confirm earlier investigations which find that TP models provide 
satisfactory fits to ROSAT and ASCA spectra of the centers of non-cooling flow clusters.  
However, TPPL models which include intrinsic absorption generally provide fits of 
superior quality, and predict that the power-law component contributes 20 percent of 
the total flux.  The photon indices of this component are typical of accreting X-ray 
sources and/or diffuse non-thermal X-ray emission which usually range from $1.0 \lesssim 
\Gamma \lesssim 3.0$.  

\subsection{Cooling Flow Clusters} 

We have fit the spectra of the CF clusters with the same variety of models used in the 
previous section, but also with an additional model; TPTPPL.  This model represents the 
emission from the observed region as the sum of the hot ICM, a diminished cooling flow 
(second TP component), and intermixed compact sources and/or extended non-thermal 
emission characterized by the PL component.

Simple TP models generally provide fairly reasonable fits to the spectra with 1.1
$\lesssim$ $\chi^{2}_{r}$ $\lesssim$ 1.25.  Derived temperatures and metallicities are  
in good agreement with previous studies (White, Jones, \& Forman 1997, Allen \& Fabian
1997; Sarazin \& McNamara 1997; Markevitch et al. 1998; Allen 2000; AB2000, Allen et al. 
2001).  None of the fits were significantly improved by letting the value of $N_H$ 
vary.  Hence, we find no evidence for excess absorption in these clusters, which is also 
consistent with several other investigations (Sarazin 1996; Allen \& Fabian 1997; 
Sarazin \& McNamara 1997; Allen 2000; AB2000). 

As anticipated, pure PL models (with or without intrinsic absorption) do not provide 
acceptable fits to the X-ray spectra of any of the CF clusters.  Typically, $\chi_r^2$ 
$\gtrsim$ 1.5 for this model.

Our TPPL models fit the spectra of all 5 CF clusters very well, with 1.0 $\lesssim$
$\chi^{2}_{r}$ $\lesssim$ 1.2 (see Table 4).  With the exception of A2029, which is best 
fit by simple TP models, the PL component typically contributes 5 - 30 \%  (mean of 20 
\%) of the total flux when absorbed only by the Galactic column.  The inclusion of 
intrinsic absorption significantly improves the quality of the fit and raises the 
contribution of the PL component to 9 - 47 \% (mean of 27 \%).  The mean  
F-statistic value for the addition of this power-law component is $\approx 50$ which 
suggests it is significant ($>$ 99 \% confidence).  The luminosity (0.7 - 10.0 keV) of 
PL components in these four clusters span $\approx 0.9 - 8 \times 10^{44}$ ergs 
s$^{-1}$, with a mean value of $\approx 3.8 
\times 10^{44}$ ergs s$^{-1}$.  The photon indices of PL components are reasonably well 
constrained and range from 1.0 $\lesssim \Gamma \lesssim$ 2.4 (mean of 1.8) (see Table 
4).  The implied intrinsic column densities acting on the PL component are rather high 
for three of the CF clusters (A644, A1689, \& A2597) and are consistent with high-mass 
compact sources.  As with the NCF clusters, a lower temperature is found for the thermal 
component than is found from the simple TP models.

TPTP models were also implemented.  The quality of the fits provided by these models is
generally comparable to that of TPPL models.  Calculation of the F-statistic 
suggests 
that the addition of the extra TP component is significant.  Thus, we confirm 
previous reports that multi-phase thermal plasmas fit the spectra of CF clusters much
better than that of isothermal models.  The temperatures and metallicites of the two 
thermal components are consistent with the results of previous studies (White, Jones, \& 
Forman 1997, Markevitch et al. 1998; Allen 2000; AB2000; Allen et al. 2001).  
For the cooler component in particular, we find $1.4 \lesssim kT \lesssim 2.2$ and 
a mean contribution of roughly 22 \% of the total flux, which is typical of cooling flow 
clusters (Peres et al. 1998).


The standard cooling flow model (TPCF) provides adequate fits to the spectra of the
CF clusters, as expected.  Derived mass deposition rates, temperatures, and metallicities  
are generally consistent with previous studies (White, Jones, \& Forman 1997; Peres et 
al. 1998; Allen et al. 1999; Allen 2000; AB2000).  The best fitting TPCF model to the 
spectrum of A644 is presented in Figure 2.  Only the ASCA SIS0 and GIS2 data are plotted.
A comparison of the TPCF and TPPL models is presented in Table 4.  It is clear that the 
TPPL models fit the spectra of the 5 CF clusters just as well as standard TPCF models 
and without any additional complexity.

\vskip0.1in
 
{\epsscale{1.0}
\plotone{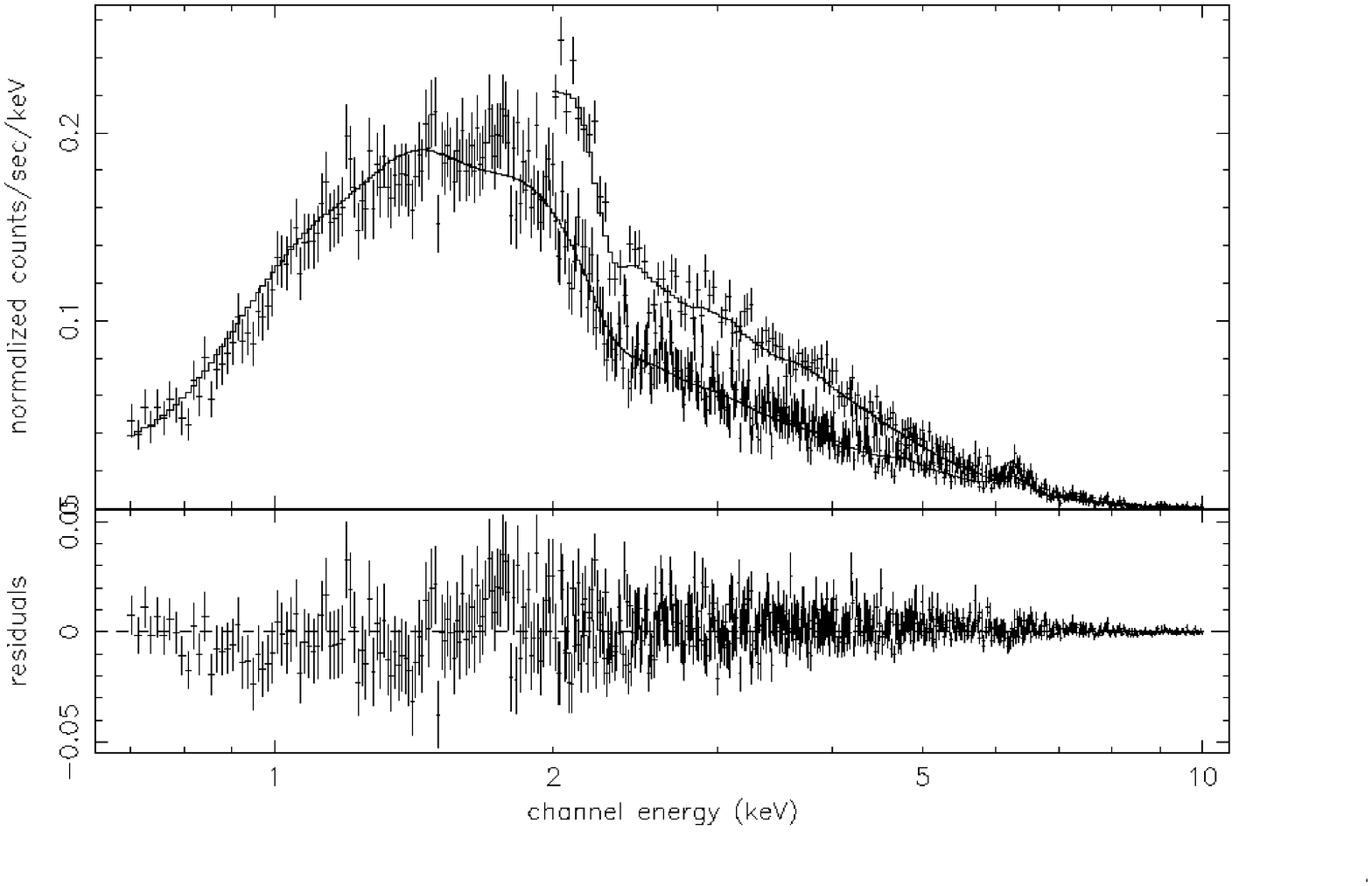}
{Fig.2. \footnotesize
 ASCA SIS0 and GIS2 spectra of the CF cluster A644.  The SIS0 spectrum is
from 2.0 - 10.0 keV while the GIS2 spectrum ranges from 0.7 - 10.0 keV.  Both data
sets are shown in the residuals in the bottom part of the plot.  The best fitting TPCF
model predicts a mass deposition rate of 200 $M_{\odot}$ yr$^{-1}$.
}}
 
\vskip0.1in

Finally, the TPTPPL model also fits the X-ray spectra of these clusters well, with 1.0
$\lesssim$ $\chi^{2}_{r}$ $\lesssim$ 1.2.  When absorbed by the Galactic column only, the 
PL component typically contributes 2 - 5 \% of the total flux inside 3 arcminutes.  The 
addition of intrinsic absorption on the PL component, again, improves the fit and the
mean contribution of the PL component is raised to $\approx$ 32 \%.  When 
compared to the TPPL model, the mean F-statistic value is about 5 ($\approx$ 68 \% 
confidence) which suggests the addition of a second TP component only improves the fit 
marginally.  The luminosity (0.7 
- 10.0 keV) of the PL component ranges from $1 - 6 \times 10^{44}$ ergs s$^{-1}$ with a 
mean value of $\approx 2.2 \times 10^{44}$ ergs s$^{-1}$.  The power-law components have 
photon indices of 1.8 $\lesssim \Gamma \lesssim$ 3.3 with a mean of 2.3.  With the 
exception of A2029 (which has virtually no contribution from the PL component), the 
intrinsic column densities on the PL component are $\gtrsim 10^{21}$ cm$^{-2}$.  The 
temperature of the cooler TP component did not change significantly from the value 
derived from TPTP models.  However, it's contribution to the flux was generally lower 
than that of TPTP models (mean $\sim$ 10 \%).

In summary, we find that in all cases the TPPL and TPTPPL models fit the ROSAT and ASCA
spectra of the five CF clusters just as well as the standard cooling flow model (TPCF) 
or the two-temperature thermal plasma (TPTP).  The PL components of these models are 
quite luminous (contributing between 27 - 32 \% of the total flux) and characterized by 
photon indices typical of compact accreting X-ray sources and/or extended non-thermal 
emission.  This, then, is the main conclusion of our work: {\it ROSAT and ASCA spectra 
can be satisfactorily modeled by attributing a significant part of the luminosity within 
the central regions of CF clusters to sources other than cooling gas.} 

\section{Discussion}

Our analysis shows that luminous (L$_{X, 0.7-10.0}$ $\sim$ 10$^{42-44}$ ergs s$^{-1}$) 
power-law components with $1.0 \lesssim \Gamma \lesssim 3.0$  might be present in the 
centers of rich clusters.  Including intrinsic absorption dramatically improves the 
quality of the fit;  in these circumstances, there is an apparent dichotomy in the 
importance of the PL components between NCF and CF clusters.  Not only is the typical PL 
component in the CF cluster more luminous by about a factor of 10 than that of its NCF 
counterpart, but it also contributes a larger fraction of the total X-ray flux within the 
central 3 arcminutes of its cluster (between 27 - 32 \% for CF clusters as opposed to 20 
\% for NCF clusters).  This may be consistent with the fact that the {\it cores} of 
CF clusters are more often observed to be active (i.e. radio loud), perhaps implying the 
presence of a central massive black hole(s) and a surrounding diffuse plasma of 
relativistic electrons, than the cores of NCF clusters (Burns 1990).  It is also worth 
noting that the improvement of the fit to the spectra of CF clusters far exceeds the 
improvement of the fit to the spectra of NCF clusters when a power-law component is 
included in the spectral model.

The most attractive feature of the TPPL models is that one does not expect large
quantities of atomic or molecular gas to be present in the centers of clusters, nor does
one expect to see emission lines below a few keV, because cooling is not significant.  
This is consistent with current observations.  On the other hand, evidence which 
apparently supports the existence of cooling flows (e.g. excess absorption, detection 
of CO line emission in extreme CF clusters) may contradict this model.  However,
the success of the single-temperature plasma + power-law model in fitting the spectra of
the central regions of rich clusters, regardless of cooling flow status, has also been
reported by a number of authors using a variety of X-ray instruments, including
Markevitch \& Vikhlinin (1997) (ASCA and ROSAT data), Markevitch et al. (1998)
(ASCA data), Rephaeli, Gruber \& Blanco (1999) (RXTE data), Guainazzi \& Molendi (1999)
(Beppo-SAX data), Fusco-Femiano et al. (1999; 2000; 2001) (Beppo-SAX data), and recently
by Sambruna et al. (2000) who used Chandra to uncover a previously {\it hidden} AGN in
Hydra A with $L_{X,2-10} \gtrsim 10^{42}$ ergs s$^{-1}$ and $\Gamma \sim 1.7$.  However,
our study is the first that systematically addresses the problems of the standard cooling 
flow model by implementing the single-temperature plasma + power-law model.

The TPTPPL models, which might represent a compromise between the standard cooling flow 
model and our favoured TPPL models, also provide good fits to the spectra of the CF 
clusters.  However, the addition of the second, cooler TP component only marginally 
improves the fit and at the expense of adding much more complexity (and
uncertainty) to the model.  Furthermore, the predicted temperatures and relative 
contributions of the cooler TP component in the TPTPPL models suggest that emission 
lines below a few keV should be present, however they have failed to turn up in Chandra 
and XMM spectra.  Finally, because the mean contribution of the cooler component is 
$\sim$ 10 \% , as compared with 20 \% in the TPTP models, this implies that mass 
deposition rates are only roughly reduced by a factor of two.  Using models similar to 
our TPTPPL models, Allen et al. (2001) reached the same conclusion for a number of 
clusters in their sample.  These reduced deposition rates are still far too large to be 
consistent with the small amounts of atomic and molecular gas found in cooling flow 
clusters.  We expect that Chandra and XMM-Newton will place much tighter constraints on 
the parameters of the complex TPTPPL models and it will be interesting to see if the 
models can be reconciled with problems of the standard cooling flow model.

Assuming that a power-law component is, indeed, present in the spectra of these clusters,
some of the possible candidates are: 

\begin{itemize}
  
\item{}Seyfert-like AGN could contribute most, if not all, of the PL component 
luminosity found in our study.  These objects can be quite luminous and are often 
observed to have $L_{X}$ $\sim 10^{41-44}$ ergs s$^{-1}$ with power-law indices of 1.6 - 
2.0 (Nandra et al. 1997).  In fact, Allen et al. 2001 have suggested that AGN are the 
likely candidate for the observed PL component in their sample.  Likewise, the 
luminosities and steepness of the PL components in our sample are generally consistent 
with emission from a central AGN(s).

\item{}Diffuse non-thermal emission, which has possibly been seen in the central 
regions of Coma, A2256, and A2199.  Searches of those clusters using Beppo-SAX and RXTE 
have yielded very luminous ($\approx$ 10$^{42-44}$ ergs s$^{-1}$) sources which are 
described by power-laws with $\Gamma \sim 0.5 - 6$ and may possibly be linked to 
power-law components observed in the ASCA band (Rephaeli, Gruber, \& Blanco 1999; 
Fusco-Femiano et al. 1999; 2000; 2001).  Unfortunately, X-ray spectra usually have to be 
extracted from regions {\it much} larger than that of the central galaxy and, thus, it is 
difficult to draw conclusions about the source of the power-law component.  A VLA survey 
of more than 200 nearby clusters found that only 29 had extended radio emission on Mpc 
scales (Giovannini, Tordi, \& Feretti 1999) perhaps implying that extended non-thermal 
emission is not that common in galaxy clusters.  It should be noted, however, that the 
detection of true intracluster radio emission is exceptionally difficult, as one must 
remove bright radio galaxies from the analysis (Rephaeli 2001).  Furthermore, the 
non-thermal emission may be associated with the radio galaxies themselves (e.g. in the 
radio lobes).

\item{}Massive black holes associated with elliptical and lenticular galaxies.  The 
regions sampled in our study generally include a single giant elliptical or cD galaxy 
plus a number of large galaxies, probably E or S0 galaxies.  Colbert \& Mushotzky (1999) 
and Allen, Di Matteo, \& Fabian (2000) have recently discovered hard power-law components 
in several nearby gE galaxies with $10^{37}\lesssim L_X \lesssim 10^{42}$ ergs s$^{-1}$ 
with photon indices of 0.5 - 3.0.  Such sources are typically less luminous than Seyfert 
nuclei, perhaps as a result of low radiative efficiency from an advection-dominated 
accretion flow (ADAF; Narayan \& Yi).  While several of these sources could, in 
principle, account for the PL component in the NCF clusters they would have to be scaled 
up by a factor of 100 to account for the emission in the CF clusters.  Faint galaxies, on 
the other hand, are much more numerous and many of them might harbor less massive central 
black holes.  The Local Group elliptical M32, whose central object has $L_X \sim 10^{37}$ 
erg s$^{-1}$ (Loewenstein et al. 1998) offers a possible guide to  the X-ray luminosity 
of faint cluster members.  Hence $\sim 10^4 - 10^6$ such  galaxies, each with a 
supermassive black hole, would be required to account for the power-law component of the 
clusters studied here.  On this basis, faint galaxies seem incapable of producing 
significant power-law emission inside the regions we have observed.  We note, however, 
that if massive black holes originally belonging to bright or faint galaxies are now 
scattered throughout intracluster space (i.e. no longer associated with optically visible 
galaxies) as the result of past mergers or stripping both of these limits may be relaxed.

\item{}Galactic X-ray binaries, having $\Gamma \approx 0 - 2.5$ (White et al. 1988, 
Verbunt et al. 1995), and which in our galaxy are most commonly associated with globular
clusters.  Intergalactic globular clusters, if similar to globular clusters in the Milky
Way, would have to be present in large numbers to collectively account for $\sim
10^{42-44}$ erg s$^{-1}$ attributable to the PL component in our models.  Assuming a
typical intergalactic globular contains a binary X-ray source with $L_X$ $\sim 10^{35}$
ergs s$^{-1}$ (Hertz \& Wood 1985) and the specific frequency of globulars in the central
gE galaxy is $S_N = 10$ (Harris 1991) then even a very bright central galaxy with M$_B$ =
-24 would have a total X-ray luminosity of $\sim 10^{39}$ ergs s$^{-1}$, too low by a
factor of $10^{4-5}$.  Even the most optimistic models of intracluster globular clusters
of West et al. (1995) predict $\sim 10^4 - 10^5$ such clusters, significantly less than is
required to account for the entire PL component.

\end{itemize}

Hence, it appears that there are probably two or three sources capable of producing the
luminosity of the PL components of our models: massive black holes associated (or once 
associated) with large elliptical galaxies, diffuse non-thermal emission, and AGN.  Does 
existing data favor one type of source over the others?  

To determine the luminosity of any point sources within the central 3 arcminutes we used 
ROSAT HRI data.  Two different procedures were implemented: I) Spectra of clearly visible 
point sources were extracted from a 5'' region centered on the source and grouped 
into one large energy bin (0.5 - 2.0 keV).  The spectra were then fit within the XSPEC 
package using a simple power-law model with $\Gamma = 1.8$ and absorbed by the Galactic 
column only.  The 0.5 - 2.0 keV flux was then converted into a 0.7 - 10.0 keV unabsorbed 
flux using the program W3PIMMS.  If more than one point source was present the 
individual fluxes were summed.  As an example, Figure 3 is the HRI observation 
of A3667, smoothed with a 5'' Gaussian beam.  There are at least two bright compact 
sources within the central 3 arcminutes of this cluster.; II) To place limits on the 
contribution of point sources which are hidden in the bright diffuse background (what we 
refer to as ``barely visible'' point sources) we calculated the mean standard deviation 
of counts per pixel (within 3') and then scaled this to a standard deviation per 
circular region of diameter = 5'' (comparable to the HRI psf).  The latter value was 
expressed as unabsorbed flux (0.7 - 10.0 keV) using the program W3PIMMS and assuming a  
simple power-law spectral model with $\Gamma = 1.8$ absorbed only by the  Galactic column 
listed in Table 1.  We regard the result as a 1-sigma limit to the contribution of any 
point sources.  

\vskip0.1in
 
{\epsscale{1.0}
\plotone{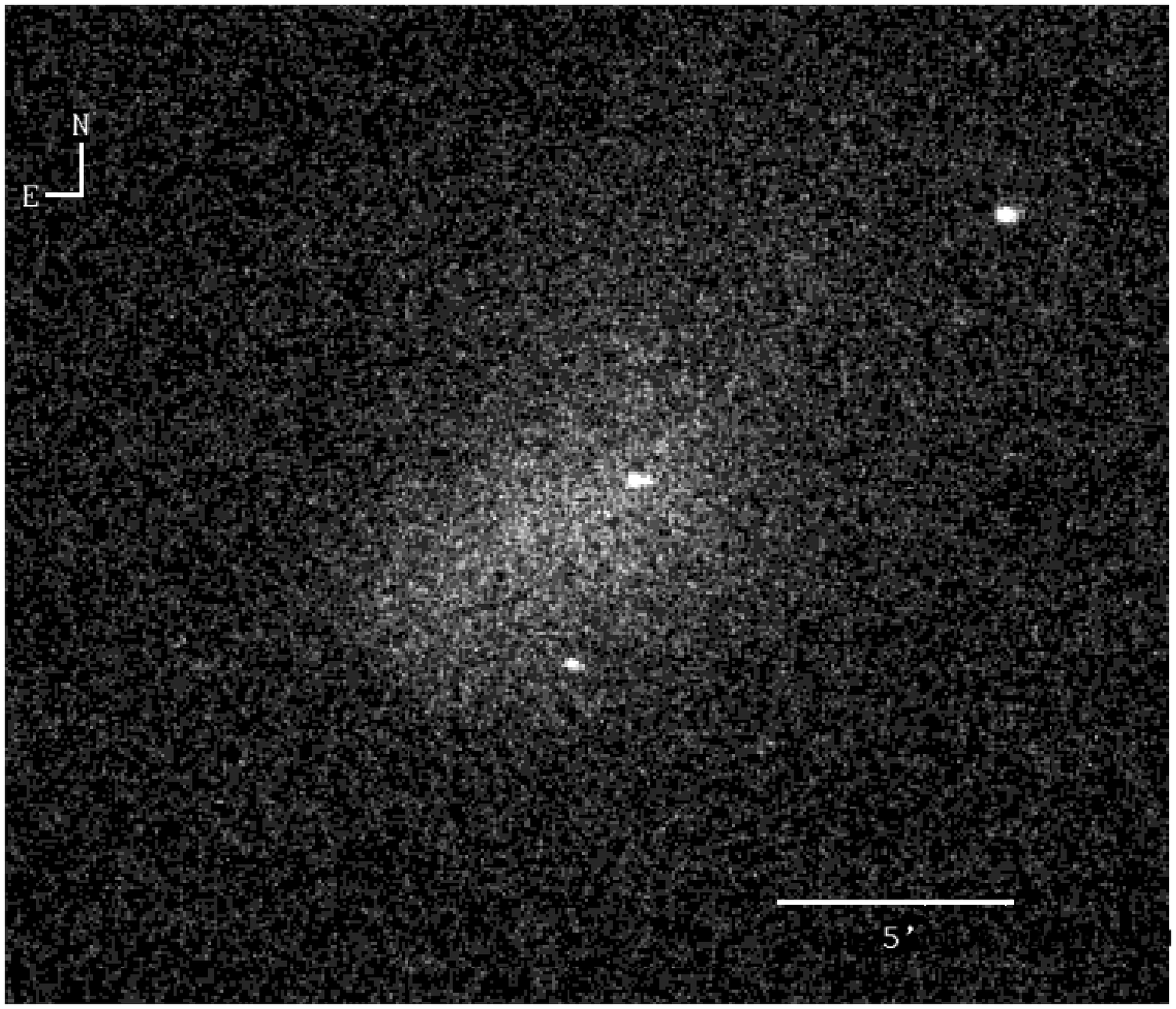}
{Fig.3. \footnotesize
ROSAT HRI observation of A3667 (centered on R.A. $20^h 12^m 31^s$, Dec.
$-56^{\circ}$49'12'', J2000).  Two extremely bright compact sources are
visible within the central 3' but together are roughly a factor of 7 less luminous then
the spectrally determined power-law component in the best fitting TPPL model.
}}
 
\vskip0.1in

The results of HRI data analysis are presented in Table 5.  One or more point sources are 
easily visible within the central 3 arcminutes of all three NCF clusters having HRI 
observations.  It is clear from a comparison with the luminosities of the 
observed PL component from the best fitting TPPL models (with intrinsic absorption), 
however, that the bright point sources cannot be solely responsible for the power-law 
emission, as they contribute only $\approx$ 5 \% of the PL component luminosity.  
Significant contributions are required from low-luminosity point sources and/or diffuse 
non-thermal emission.  The third column in Table 5 (Method II) presents the 3-sigma 
limit of ``barely visible'' point sources, suggesting that at least $10^{2-3}$ of them 
are needed (in addition to the visible point sources) to account for the entire 
power-law component.  

There are no visible point sources within the central 3' of the 5 CF clusters studied. 
This is almost certainly the result of the CF clusters being much more centrally peaked 
than the NCF clusters.  At least $10^{2-3}$ ``barely visible'' point sources 
are required to account for the observed PL component in these clusters.  

Perhaps another way to differentiate between compact sources and diffuse non-thermal 
emission is through the presence or absence of flux variability, since most luminous 
accreting X-ray sources are known to vary significantly on short time scales (e.g. 
Nandra et al. 1997).  However, because the PL component typically contributes only 
$\approx 20$ \% this might make detection of variability difficult in clusters 
(especially in the CF clusters).  Furthermore, if the PL component is produced by 
$10^{2-3}$ point sources, and the variations in each are random, it will make the 
variation in the collective luminosity quite small, perhaps rendering this method 
unusable. 

If compact or diffuse non-thermal sources provide significant emission from the centers 
of rich clusters, it will be necessary to revise a number of conclusions derived from 
X-ray data.  Fully exploring the consequences is beyond the scope of the present paper; 
however, we will mention a few important issues.  Allen (2000) reports that a 
deprojection analysis of ROSAT HRI data yields cooling flow mass deposition rates that
agree reasonably well with those derived from ASCA spectra.  A consistent re-analysis
would require first subtracting the power-law contribution from the X-ray images, but 
of course the spatial distribution of the power-law emission must first be 
understood.  The presence of emission lines from heavy ions originally led astronomers to 
accept thermal emission as the dominant mechanism in rich cluster ICMs.  Accretion-driven 
X-ray sources do exhibit line emission (Kallman, Vrtilek, \& Kahn 1989; Narayan \& 
Raymond 1999), but the lines are produced under quite different physical conditions than 
those within the canonical hot ICM.  The implications for ICM abundance estimates and 
enrichment scenarios will require careful study.  Likewise, previously derived 
temperatures of the ICM will also be affected (as demonstrated in Table 3).  Might 
emission from compact sources account for the excess hard X-radiation recently 
discovered (Xu et al. 1998; Ikebe et al. 1997) in the centers of several clusters?  The 
next generation of X-ray detectors and satellites (i.e. XMM and Chandra) will be needed 
to map the X-ray emission with adequate spatial resolution to address this and other 
questions raised by the present work.   

\section{Conclusions}

We have shown that ROSAT PSPCB and ASCA GIS2/GIS3 spectra of ten X-ray luminous
clusters permit a significant contribution from compact sources and/or diffuse 
non-thermal emission.  Although this conclusion holds irrespective of the presence of 
a cooling flow, we find that power-law components are especially luminous in clusters 
with massive cooling flows.  Our study indicates that inferred cooling flow mass 
deposition rates may be lowered substantially, thus ameliorating, or possibly resolving,  
the long-standing problem of the missing cool gas.  Existing X-ray 
observations of large elliptical galaxies, AGN, and extended non-thermal emission 
suggest that any of these sources could account for the observed luminosity of the 
power-law component.  However, the absence of extremely bright point sources in the 
ROSAT HRI observations rules out the possibility that the power-law emission comes from 
a few extremely bright compact sources.  More likely is that most of the power-law 
emission is produced by some combination of diffuse and low-luminosity compact sources, 
such as massive black holes.  The deep imaging capabilities and  superior spectral 
resolution of Chandra and XMM should be able to place much tighter constraints on the 
relative contributions of power-law sources and thermal bremsstrahlung to cluster X-ray 
emission and, perhaps, even determine the source of the power-law emission.   

\noindent{\bf Acknowledgments}

We thank the ROSAT and ASCA help teams, in particular Jane Turner, for advice with data
reduction techniques.  We also thank Paul Nandra, Peter Thomas, Arif Babul, Ann Gower, 
Pat Henry, and Alastair Edge for useful comments and suggestions.  IGM acknowledges 
support from NSERC of Canada (USRA \& PGSA).  GAW and MJW were supported by NSERC 
Research grants.  MJW also acknowledges support from NSF grant AST-0071149.

\newpage

\begin{deluxetable}{ccccccll}
\tablecaption{Clusters investigated in this study.
\label{data}}
\tablehead{
\colhead{Cluster}                      &
\colhead{$z$ \tablenotemark{a}}        &
\colhead{$N_H$ \tablenotemark{b}}    &
\colhead{Type }       &
\colhead{$\dot{M}$ \tablenotemark{c}}     &
\colhead{kT  \tablenotemark{d}}     &
\colhead{Notes}   &
}
\tablenotetext{a}{Redshifts from the NASA/IPAC Extragalactic database}
\tablenotetext{b}{Galactic HI column density (in units of $10^{20}$ cm$^{-2}$)  from 
Dickey \& Lockman (1990), except where noted.}
\tablenotetext{c}{Mass deposition rate (in $M_{\odot}$ yr$^{-1}$) from Peres et al. 
(1998)}
\tablenotetext{d}{Single-phase ICM temperature (in keV) reported by White, Jones, \& 
Forman (1997).}
\tablenotetext{e}{Galactic HI column density (in units of $10^{20}$ cm$^{-2}$) from 
Stark et al. (1992).}
\tablenotetext{f}{Feretti et al. (1999)}
\tablenotetext{g}{Bauer \& Sarazin (2000)}
\tablenotetext{h}{Molendi \& De Grandi (1999)}
\tablenotetext{i}{Oegerle \& Hill (1994)}
\tablenotetext{j}{Markevitch et al. (2000)}
\tablenotetext{k}{Bazzano et al. (1984)}
\tablenotetext{l}{Sarazin et al. (1995)}
\tablenotetext{m}{Flores, Quintana, \& Way (2000)}
\tablenotetext{n}{Bardelli, Zucca, \& Baldi (2001)}
\tablenotetext{o}{Fusco-Femiano et al. (2001)}
\startdata
A119  & 0.0442  & 3.17  & NCF & 0   & 5.3  & 3 bright radio galaxies near center$^f$\\
A644  & 0.0704  & 8.00$^e$ & CF  & 189 & 6.5  & minor merger event$^g$\\
A1689 & 0.1832  & 1.81  & CF  & 645 & 8.7 & \nodata \\
A2029 & 0.0773  & 3.14  & CF  & 556 & 7.4  & non-thermal emission not detected$^h$\\
A2063 & 0.0353  & 2.90$^e$ & NCF  & 37  & 4.1 & subcluster merger?$^i$\\
A2142 & 0.0909  & 4.16  & CF  & 350 & 11.4 & merger$^j$ \& hard X-ray emission 
detected$^k$\\
A2597 & 0.0852  & 2.49  & CF  & 271 & 9.1 & powerful cD radio galaxy$^l$ \\
A3266 & 0.0589  & 3.00$^e$ & NCF & 0   & 6.2 & major merger$^m$ \\
A3532 & 0.0554  & 5.98  & NCF & 0   & 4.7 & merger$^n$ \\
A3667 & 0.0556  & 4.00$^e$ & NCF & 0   & 7.1  & hard X-ray emission found$^o$\\
\enddata
\end{deluxetable}

\newpage

\begin{deluxetable}{ccccccccc}
\tablecaption{Comparison with AB2000.
\label{AB2000}}
\tablewidth{38pc}
\tablehead{
\colhead{Cluster}                      &
\colhead{Model}        &
\colhead{$kT_{cool}$ \tablenotemark{a}}    &
\colhead{$\dot{M}$ \tablenotemark{b} }       &
\colhead{$\chi_r^2$ } &
\colhead{}     &
\colhead{$kT_{cool}$ \tablenotemark{c}}    &
\colhead{$\dot{M}$ \tablenotemark{d} }       &
\colhead{$\chi_r^2$ } }
\tablenotetext{a}{Temperature (in keV) of the cooler TP component found by AB2000.}
\tablenotetext{b}{Mass deposition rate (in $M_{\odot}$ yr$^{-1}$) found by AB2000.}
\tablenotetext{c}{Temperature (in keV) of the cooler TP component found in this study.}
\tablenotetext{d}{Mass deposition rate (in $M_{\odot}$ yr$^{-1}$) found in this study.} 
\startdata

A119 & TP & - & - & 0.987 & & - & - & 1.011 \\& TPCF & 0.08 $\pm$ $\infty$ & 
0 $\pm$ 3 & 0.994 &  & 0.01 $\pm$ 151 & 0.6 $\pm$ 5 & 1.026 \\
A2029 & TPTP & 1.29 $\pm$ 0.15 & - & 1.592 &  & 1.19 $\pm$ 0.31 & - & 1.499 \\
& TPCF & 0.08 $\pm$ 6.97 & 132 $\pm$ 31 & 1.911 &  & 0.03 $\pm$ 4.82 & 170 $\pm$ 43 & 
1.772 \\
A2142 & TPCF & 0.08 $\pm$ 7.41 & 188 $\pm$ 43 & 1.190 &  & 0.10 $\pm$ 9.16 & 230
$\pm$ 73 & 1.130 \\ 
\enddata
\end{deluxetable}

\newpage

\begin{deluxetable}{cccccccc}
\tablecaption{Comparison of the TP and TPPL models for the NCF clusters.
\label{NCF}}
\tablewidth{36pc}
\tablehead{
\colhead{Cluster}                      &
\colhead{Model}        &
\colhead{kT \tablenotemark{a}}    &
\colhead{$\Gamma$ \tablenotemark{b} }       &
\colhead{PL \% \tablenotemark{c}}     &
\colhead{$N_{H,PL}$  \tablenotemark{d}}     &
\colhead{$\chi^{2}/d.o.f.$}   &
}
\tablenotetext{a}{Temperature (in keV) of the thermal plasma component.}
\tablenotetext{b}{Photon index of the power-law component.}
\tablenotetext{c}{Contribution of the PL component to the total flux (0.14 - 
10.0 keV).}
\tablenotetext{d}{Intrinsic HI column density (in units of $10^{20}$ cm$^{-2}$) acting on 
PL component. }
\startdata

A119 & TP & $5.91 \pm 0.20$ &\nodata & \nodata & \nodata & 467.4/446 \\
 & TPPL& $3.96 \pm 0.78$ & $1.55 \pm 0.17$ & 25 \% & $29.5 \pm 17.6$ & 454.0/443\\ 
A2063 & TP & $3.09 \pm 0.05$ & \nodata & \nodata & \nodata& 945.6/807 \\
 & TPPL & $2.49 \pm 0.23$  & $1.41 \pm 0.27$ & 8.5 \% & $47.5 \pm 71.5$ & 933.3/804\\
A3266 & TP & $8.48 \pm 0.24$ & \nodata & \nodata & \nodata& 1160.5/1117 \\
 & TPPL & $4.23 \pm 0.86$ & $1.33 \pm 0.12$ & 23 \% & $140.0 \pm 35.2$&  1158.0/1114\\
A3532 & TP & $4.55 \pm 0.13$ & \nodata & \nodata & \nodata & 776.4/746\\
 & TPPL & $ 3.71 \pm 1.14$ & $2.12 \pm 0.10$ & 87.5 \% & $18.0 \pm 2.70$ & 756.0/743\\
A3667 & TP & $5.64 \pm 0.14$ &  \nodata & \nodata & \nodata& 809.0/791 \\
 & TPPL & $3.22 \pm 0.32$ &  $1.30 \pm 0.15$ & 22 \% & $88.0 \pm 40.4$ & 797.6/788\\

\enddata
\end{deluxetable}

\newpage
 
\begin{deluxetable}{cccccccc}
\tablecaption{ Comparison of the TPPL and TPCF models for the CF clusters.
\label{CF}}
\tablewidth{36pc}
\tablehead{
\colhead{Cluster}                      &
\colhead{Model}        &
\colhead{$\dot{M}$ \tablenotemark{a}}  &
\colhead{$\Gamma$ \tablenotemark{b} }       &
\colhead{PL \% \tablenotemark{c}}     &
\colhead{$N_{H,PL}$  \tablenotemark{d}}     &
\colhead{$\chi^{2}/d.o.f.$}   &
}
\tablenotetext{a}{Mass deposition rate (in $M_{\odot}$ yr$^{-1}$).}
\tablenotetext{b}{Photon index of the power-law component.}
\tablenotetext{c}{Contribution of the PL component to the total flux (0.14 -
10.0 keV).}
\tablenotetext{d}{Intrinsic HI column density (in units of $10^{20}$ cm$^{-2}$) acting on 
PL component. }
\startdata

A644 & TPPL & \nodata & $2.37 \pm 0.78$  & 35 \%  & $24.1 \pm 20.6$ & 1855.7/1613 \\
 & TPCF   & $200 \pm 71$ &\nodata  & \nodata  & \nodata & 1841.2/1613 \\
A1689 & TPPL & \nodata& $1.02 \pm 0.29$ & 9 \% & $166.0 \pm 71.5$ & 1643.6/1409 \\
 & TPCF & $504 \pm 149$ & \nodata & \nodata & \nodata & 1653.9/1409\\
A2029 & TPPL & \nodata & $2.13 \pm 1.44$ & 5 \% & $3.34 \pm 6.40$ & 1460.3/1149 \\
 & TPCF & $427 \pm 60$ & \nodata & \nodata & \nodata & 1462.1/1149\\
A2142 & TPPL & \nodata & $1.67 \pm 0.09$ & 47 \% & $2.14 \pm 0.54$ & 1010.4/868\\
 & TPCF & $307 \pm 90$& \nodata & \nodata & \nodata  & 1013.2/868\\
A2597 & TPPL & \nodata& $1.89 \pm 0.45$ & 17 \% & $177.0 \pm 72.0$ & 1305.1/1281  \\
 & TPCF & $217 \pm 23$ & \nodata & \nodata & \nodata & 1302.3/1281\\

\enddata
\end{deluxetable} 

\newpage
 
\begin{deluxetable}{ccccc}
\tablecaption{Limits in cgs units on point source contribution from HRI data.
\label{HRI}}
\tablewidth{25pc}
\tablehead{
\colhead{Cluster}                      &
\colhead{Method I \tablenotemark{a}}        &
\colhead{Method II \tablenotemark{b}}  &
\colhead{$L_{X,PL}$ \tablenotemark{c}} &
}
\tablenotetext{a}{Luminosity  of visible point sources.}
\tablenotetext{b}{Upper limit of luminosity  of ``barely visible'' 
point sources within the HRI psf.}
\tablenotetext{c}{Luminosity of the PL component of the best fitting TPPL 
models with intrinsic absorption.}

\startdata
A119 & $8.12 \times 10^{41}$ & $\leq 5.40 \times 10^{40}$ & $1.67_{-0.97}^{+1.46} 
\times 10^{43}$\\
A644 & \nodata  & $\leq 2.99 \times 10^{41}$ & $2.59_{-1.42}^{+2.74} \times 10^{44}$\\
A1689 & \nodata & $\leq 2.09 \times 10^{42}$ & $2.95_{-2.95}^{+4.70} \times 10^{44}$\\
A2029 & \nodata & $\leq 5.34 \times 10^{41}$ & $8.65_{-8.65}^{+31.68} \times 10^{43}$\\
A2142 & \nodata & $\leq 6.45 \times 10^{41}$ & $8.81_{-2.95}^{+4.22} \times 10^{44}$\\
A2597 & \nodata & $\leq 7.02 \times 10^{41}$ & $0.94_{-0.94}^{+2.37} \times 10^{44}$\\  
A3266 & $1.47 \times 10^{42}$ & $\leq 8.40 \times 10^{40}$ & $6.28_{-5.11}^{+7.91} \times 
10^{43}$ \\ 
A3667 & $8.01 \times 10^{42}$ & $\leq 8.97 \times 10^{40}$ & $5.70_{-3.89}^{+5.46} \times 
10^{43}$\\
\enddata
\end{deluxetable}

\end{document}